\documentclass[aps, prd, onecolumn, tightenlines, notitlepage, superscriptaddress, nofootinbib, preprintnumbers, floatfix,showkeys,11pt]{revtex4-2}

\usepackage[normalem]{ulem}
\usepackage{amstext}
\usepackage{amssymb}
\usepackage{amsmath}
\usepackage{graphicx}
\usepackage{url}
\usepackage{color}
\usepackage{ulem}
\usepackage[utf8]{inputenc}
\pdfoutput=1
\usepackage{textcomp}
\usepackage{booktabs,siunitx,array,threeparttable}
\usepackage{gensymb}

\usepackage{epsfig,amsfonts,mathrsfs,amsmath,amssymb,graphicx,color,slashed,multirow}

\usepackage[x11names]{xcolor}
\usepackage[colorlinks]{hyperref}

\usepackage{textgreek} 
\usepackage{caption, subcaption} 
\usepackage{booktabs} 
\usepackage{physics}

\definecolor{lightapricot}{rgb}{0.99, 0.84, 0.69}
\definecolor{nicered}{rgb}{0.7,0.1,0.1}
\definecolor{nicegreen}{rgb}{0.1,0.5,0.1}
\definecolor{coral}{rgb}{1.0, 0.5, 0.31}
\definecolor{blue(ncs)}{rgb}{0.0, 0.53, 0.74}
\definecolor{darkspringgreen}{rgb}{0.09, 0.45, 0.27}
\definecolor{seagreen}{rgb}{0.18, 0.55, 0.34}

\definecolor{cadmiumgreen}{rgb}{0.0, 0.42, 0.24}
\definecolor{chromeyellow}{rgb}{1.0, 0.65, 0.0}
\definecolor{darkturquoise}{rgb}{0.0, 0.81, 0.82}
\definecolor{denim}{rgb}{0.08, 0.38, 0.74}
\definecolor{purple(x11)}{rgb}{0.63, 0.36, 0.94}
\definecolor{red(ncs)}{rgb}{0.77, 0.01, 0.2}
\definecolor{ruddypink}{rgb}{0.88, 0.56, 0.59}
\definecolor{slateblue}{rgb}{0.42, 0.35, 0.8}
\definecolor{airforceblue}{rgb}{0.36, 0.54, 0.66}
\definecolor{orange(colorwheel)}{rgb}{1.0, 0.5, 0.0}

\makeatletter
    \newcommand{\colorboxed}[3][white]{\fcolorbox{#2}{#1}{\m@th$\displaystyle#3$}}
\makeatother
\AtBeginDocument{\hypersetup{citecolor=seagreen,linkcolor=seagreen,urlcolor=seagreen}}
\usepackage{appendix}

\begin{document}
\preprint{IFIC 24/17}
\title{{\LARGE Quantum Decoherence effects on precision measurements at DUNE and T2HK}}

\author{G. Barenboim}
\email{gabriela.barenboim@uv.es}
\affiliation{Instituto de Física Corpuscular, CSIC-Universitat de València, C/Catedrático José Beltrán 2, Paterna 46980, Spain}
\affiliation{Departament de Física Teòrica, Universitat de València,
C/ Dr. Moliner 50, Burjassot 46100, Spain}

\author{A. Calatayud-Cadenillas}
\email{anthony.calatayud@pucp.edu.pe}
\affiliation{Seccíon Física, Departamento de Ciencias, Pontificia Universidad Cat\'olica del Per\'u, Apartado 1761, Lima, Per\'u}

\author{A. M. Gago}
\email{agago@pucp.edu.pe}
\affiliation{Seccíon Física, Departamento de Ciencias, Pontificia Universidad Cat\'olica del Per\'u, Apartado 1761, Lima, Per\'u}

\author{C. A. Ternes}
\email{christoph.ternes@lngs.infn.it}
\affiliation{Istituto Nazionale di Fisica Nucleare (INFN), Laboratori Nazionali del Gran Sasso, 67100 Assergi, L’Aquila (AQ), Italy
}

\begin{abstract}
We investigate the potential impact of neutrino quantum decoherence on the precision measurements of standard neutrino oscillation parameters in the DUNE and T2HK experiments. We show that the measurement of $\delta_\text{CP}$, $\sin^2\theta_{13}$ and $\sin^2\theta_{23}$ is stronger effected in DUNE than in T2HK. On the other hand, DUNE would have a better sensitivity than T2HK to observe decoherence effects. By performing a combined analysis of DUNE and T2HK we show that a robust measurement of standard parameters would be possible, which is not guaranteed with DUNE data alone.
\end{abstract}

\maketitle

\section{Introduction}
Quantum field theory and general relativity stand as remarkably successful theories, each validated through precise experimental confirmation. Nevertheless, a fundamental incompatibility exists between these two theories: quantum wave equations are formulated within a fixed space-time framework, whereas general relativity posits a dynamic space-time that curves in response to the distribution of matter. Ultimately, general relativity proves non-renormalizable when approached as a quantum field theory. 

Numerous efforts to reconcile this tension have concentrated on quantizing gravity, and the primary contenders are string theory and loop quantum gravity. However, a definitive consensus on which will ultimately yield the fundamental theory is yet to be reached.
These approaches, which can be termed the 'top-down' method, entail formulating a theory and subsequently assessing whether it can generate testable physical predictions.
An alternative approach, the one advocated here, is the 'bottom-up' method. In this method, experimental data is analyzed, and various phenomenological models are tested to fit the data. 

Quantum gravity holds the potential to introduce significant alterations to fundamental physics, and the ultimate theory is likely to encompass unforeseen effects on basic physical principles. Our focus in this context is solely on one of these mechanisms: the impact of quantum gravity on the standard quantum mechanical time-evolution.

At first glance, one might anticipate that experimental physics would be incapable of probing quantum gravity, given that its pertinent energy scale is the Planck energy. However, as we will see, neutrinos are the ideal tool for such  an exploration.
Due to the small mass differences between neutrinos in comparison to the long coherence lengths for neutrino oscillations,  microscopic quantum effects become observable over macroscopic distances and any distortion gets equally amplified.

In the realm of quantum gravity, space-time is anticipated to exhibit a 'foamy' structure, where minuscule black holes emerge from the vacuum, akin to the formation of quantum particles. These tiny black holes are ephemeral, swiftly dissipating through Hawking radiation. Each microscopic black hole has the potential to introduce fluctuations in the standard time-evolution of quantum mechanics. Initially, a pure quantum state describing the space-time prior to black hole formation may evolve into a mixed quantum state, characterizing the residual Hawking radiation after the black hole's evaporation. Notably, in the domain of quantum gravity, the conventional notion that pure states cannot transition into mixed states is no longer an absolute rule. 

In order to describe this process, the density matrix formalism needs to be introduced. To  facilitate the transition from pure to mixed states,  one needs to  modify the differential equation that describes the time-evolution of the density matrix, $\rho$. Several conditions must be imposed to guarantee the conservation of probability and to ensure that the trace of $\rho^2$ never exceeds unity.

Alterations of quantum mechanics, whether arising from quantum gravity or other sources, are frequently referred to as quantum decoherence effects. These effects arise due to dissipative interactions with an environment, with space-time foam being just one example. They enable transitions from pure to mixed quantum states, signifying a loss of coherence \cite{Lindblad:1975kmh, Gorini:1975nb}

The loss of coherence does not prevent neutrino flavor oscillations, but modifies the oscillation phenomenon through distance- (and energy-dependent) damping terms. This might suggest that long-baseline experiments are better suited for probing neutrino quantum decoherence. 
Solar neutrinos have been used to place very strong limits on energy-independent quantum decoherence, as shown in \cite{Fogli:2007tx,deHolanda:2019tuf}, in combination with reactor antineutrino data from the Kamioka Liquid Scintillator Antineutrino Detector (KamLAND). Note, however, that depending on the model of decoherence (i.e. in the absence of relaxation terms), all sensitivity to measure decoherence parameters might be lost if neutrino oscillations become averaged, as in the case of solar neutrinos. 
In Refs.~\cite{Lisi:2000zt,DEsposito:2023psn} atmospheric neutrinos are used to bound neutrino quantum decoherence using Super-Kamiokande data. Currently, the most stringent limits on decoherence parameters with positive\footnote{If the origin of the decoherence effects is related to quantum gravity~\cite{Barenboim:2004wu,Barenboim:2006xt,Benatti:2001fa,Mavromatos:2006yn}, one would naturally expect $n\geq0$.} energy dependence ($\Gamma\propto E^{n}, n>0$, see below) have been obtained using atmospheric neutrinos observed at the IceCube Neutrino Observatory~\cite{IceCube:2023gzt,Coloma:2018idr}.
Data from long-baseline accelerator and reactor experiments have also been used to set limits on decoherence parameters~\cite{Blennow:2005yk,deOliveira:2013dia,BalieiroGomes:2016ykp}. Further analyses include accelerator data from MINOS and Tokai to Kamioka (T2K)~\cite{Gomes:2020muc}, as well as
from NuMI Off-Axis $\nu_e$ Appearance (NOvA) and T2K~\cite{Coelho:2017zes}. The sensitivity at the Deep Underground Neutrino Experiment (DUNE) in and for the Jiangmen Underground Neutrino Observatory (JUNO) have been presented in Refs.~\cite{BalieiroGomes:2018gtd,Carpio:2018gum,JUNO:2021ydg}. Bounds from up-to-date data from accelerator and reactor experiments and sensitivities for DUNE and JUNO for many decoherence models with different energy dependences have been reported in Ref.~\cite{DeRomeri:2023dht}. Interestingly, neutrino quantum decoherence has been suggested as a solution of the Gallium anomaly~\cite{Farzan:2023fqa}, which is not in tension with data from other experiments~\cite{Giunti:2023kyo}, particularly from reactor rate measurements~\cite{Giunti:2021kab,Giunti:2022btk}. Indeed, the phenomenology of neutrino quantum decoherence is highly diverse, with the potential for exhibiting explicit CPT violation effects or the activation of the Majorana CP phases within the neutrino oscillation framework~\cite{Carrasco:2018sca,Carrasco-Martinez:2020mlg,Capolupo:2018hrp}.

The next generation of experiments offers an excellent opportunity to further explore these decoherence effects. Even more, it is imperative to study whether the present of these unexpected physics may be confused with or jeopardize the search for more standard physics like CP violation in the leptonic sector. In this paper, we discuss the impact decoherence effects might have on the measurements of standard parameters at DUNE and T2HK.

Our paper is structured as follows: In Section~\ref{sec:theo} we discuss the decoherence model under consideration and derive new results concerning the neutrino oscillation probability in presence of decoherence. In Section~\ref{sec:sim} we detail our simulation strategy of DUNE and T2HK. Our results are presented in Section~\ref{sec:res} and we draw our conclusions in Section~\ref{sec:conc}.

\section{Theoretical considerations}
\label{sec:theo}
The evolution of a given system formulated, either through the Schrödinger equation for pure states or the Liouville equation for mixed ones, is a distinctive feature of systems isolated from their environment. This evolution  is normally called Hamiltonian evolution and is given by
\begin{equation}
\frac{d\rho(t)}{dt} = -i \left[H,\rho(t)\right]\,,
\end{equation}
where $H$ is the Hamiltonian of the system and 
\[ \rho = \sum_i p_i \ket{\nu_i}  \bra{\nu_i} . \] If the system is not closed or the evolution is irreversible, the evolution equation incorporates a dissipative term and is called the Lindblad equation~\cite{Lindblad:1975ef},
\begin{equation}
\frac{d\rho(t)}{dt} = -i \left[H,\rho(t)\right] + L\left[\rho(t)\right]
\label{lindblad}
\end{equation}
with~\cite{Gago:2002na,Benatti:2000ph,Lisi:2000zt,Ohlsson:2000mj,Stuttard:2020qfv}
\begin{equation}
L\left[\rho(t)\right] = -\frac{1}{2} \sum_{j} \left\lbrace A^{\dagger}_{j} A_{j} \rho(t) +
\rho(t) A^{\dagger}_{j} A_{j} \right\rbrace
+\sum_{j} A_{j} \rho(t) A^{\dagger}_{j}\,.
\end{equation}
Here, $\left\{ A_j\right\} $ represents a set of dissipative operators with $ j=1,2,...,n^2-1$  being $n$ the number of neutrino flavor/mass eigenstates. The inclusion of these operators introduces a non-unitary evolution to $\rho$.  
The dissipative term has to meet the conditions of complete positivity and a time-increasing von Neumann entropy~\cite{Lisi:2000zt,Benatti:2000ph,Gago:2002na,Oliveira:2010zzd,BalieiroGomes:2018gtd,Stuttard:2020qfv,Buoninfante:2020iyr,Benatti:1987dz}; the latter is met by demanding that $A_j$ be Hermitian.
Once these conditions are imposed, Eq.~(\ref{lindblad}) can be simplified to
\begin{eqnarray}
    \dot{\rho}_0 &=&0 \nonumber \\
     \dot{\rho}_{i} &=& \sum_j M_{ij} \rho_j + M^{ji}_D \rho_j   \nonumber
\end{eqnarray}
with $\rho = \sum_i \rho_i \lambda_i$ being $\lambda_i$ the Gell-Mann matrices in the canonical order and 
$M_{kj}= \sum_i H_i f_{ijk}$ with $H$ also decomposed in the same way as $\rho$ and
$f_{ijk}$ the structure constants of the $SU(n)$ group~\cite{Benatti:2000ph,Gago:2002na,Carrasco:2018sca,Buoninfante:2020iyr,Stuttard:2020qfv}.
The matrix $M_D$ is the decoherence matrix, which is symmetric and semi-positive defined.
Thus, 
\begin{eqnarray}
    \varrho (t) = e^{(M+M_D)t} \varrho(0)
\end{eqnarray}
where $\varrho$ is an eight-dimensional column vector with the $\rho_i$ as its entries. 
Therefore, the neutrino oscillation probability can be obtained from \cite{Benatti:2000ph}
\begin{eqnarray}
    P_{\nu_{\alpha}\rightarrow \nu_{\beta}}(t) = \frac{1}{n} + \frac{1}{2} (\varrho_\beta(0))^T \varrho_\alpha (t)\,. 
    \label{ProbRhomatrix}
\end{eqnarray} 
where $\varrho_{\alpha(\beta)}$ corresponds to the $\alpha(\beta)$ flavor.

\subsection{Neutrino Oscillation Probability with Quantum Decoherence effects}
The decoherence matrix that we will use in this work obeys the following structure:
\begin{equation}
M_D^V=-\text{Diag}(\Gamma_1,\Gamma_2,\Gamma_1,\Gamma_1,\Gamma_2,\Gamma_1,\Gamma_2,\Gamma_1).
\label{diagvalid0}
\end{equation}
The reason for such a choice is not only simplicity, but mainly the fact that this matrix has the same form also in a medium of constant matter density~\cite{Carpio:2017nui}. Recall that the decoherence matrix, $M_D^V$, is defined in the vacuum basis. Given the conditions for the decoherence entries in the three-neutrino generation framework  fully described in~\cite{Carrasco:2018sca}, we have that $\frac{1}{3} \Gamma_1 \leq \Gamma_2 \leq \frac{5}{3}\Gamma_1$~\cite{Carpio:2017nui}. Therefore, we get for the three-neutrino oscillation probability in vacuum substituting the aforementioned matrix 
into Eq.~(\ref{ProbRhomatrix})\footnote{ A similar formula was obtained in Ref.~\cite{Gago:2002na}.}: 
\begin{eqnarray}
    P_{\nu_{\alpha}\rightarrow \nu_{\beta}}(t) =
    \frac{1}{3}&+&\frac{1}{2}\left\{\left[
    \sum_{ij(kl)} A^{\alpha\beta}_{ij} \cosh \left(\frac{\Omega_{kl}t}{2}\right) 
    +\left( -2 \Delta_{kl} B^{\alpha\beta}_{ij} 
    + \Delta\Gamma_{12}  C^{\alpha\beta}_{ij}\right) \frac{\sinh{\left(\frac{\Omega_{kl}t}{2}\right)}}
    {\Omega_{kl}}
    \right] e^{-\frac{1}{2}\Sigma_{12} t} \right. \nonumber \\  
       &+& e^{-\Gamma_{1}t}\rho_{3}^{\alpha}\rho_{3}^{\beta}+e^{-\Gamma_{1}t}\rho_{8}^{\alpha}\rho_{8}^{\beta}
    \Bigg\}  
    \label{Probgral}
\end{eqnarray}
 where $ij(kl)=12(21),45(31),67(32)$, $\Delta\Gamma_{12}=\Gamma_{2}-\Gamma_{1}$, $\Sigma_{12}=\Gamma_{2}+\Gamma_{1}$, $\Omega_{kl}=\sqrt{\Delta\Gamma^2_{12}-4\Delta_{kl}^{2}}$ and $\Delta_{kl}=\frac{\Delta m_{kl}^{2}}{2 E}$, with 
 $\Delta m_{kl}^{2}=m_k^2-m_l^2 $, ($k,l=1,2,3$). Here $A^{\alpha\beta}_{ij}$, $B^{\alpha\beta}_{ij}$, and $C^{\alpha\beta}_{ij}$ are defined as
 \begin{eqnarray}
  A^{\alpha\beta}_{ij} &=& \left( \rho_{i}^{\alpha}\rho_{i}^{\beta}+\rho_{j}^{\alpha}\rho_{j}^{\beta}\right)\,, \\
  B^{\alpha\beta}_{ij} &=& \left( \rho_{i}^{\alpha}\rho_{j}^{\beta}-\rho_{j}^{\alpha}\rho_{i}^{\beta}\right)\,, \\
 C^{\alpha\beta}_{ij} &=& \left( \rho_{i}^{\alpha}\rho_{i}^{\beta}-\rho_{j}^{\alpha}\rho_{j}^{\beta}\right)\,.    
\end{eqnarray} 
Using the definitions for $\rho_{\mu}^{\alpha} (\mu=0,..,8)$ given in Appendix~\ref{sec:AppendixA} we can write $A^{\alpha\beta}_{ij}$, $B^{\alpha\beta}_{ij}$, and $C^{\alpha\beta}_{ij}$ in 
a more familiar fashion $(k>l)$: 
\begin{eqnarray}
  A^{\alpha\beta}_{ij} &=& 4 \, \mathfrak{Re}\left( U^{*}_{\alpha l}
    U_{\alpha k}
    U_{\beta l}
    U^{*}_{\beta k}
    \right)\\
  B^{\alpha\beta}_{ij} &=& 4 \,\mathfrak{Im}
 \left( U^{*}_{\alpha l}
    U_{\alpha k}
    U_{\beta l}
    U^{*}_{\beta k}
    \right)  \\
 C^{\alpha\beta}_{ij} &=& 4 \,
 \mathfrak{Re} \left( U^{*}_{\alpha l}
    U_{\alpha k}
    U^{*}_{\beta l}
    U_{\beta k}  \right)
  \label{coeff}
\end{eqnarray}
Similarly
\begin{eqnarray}
\rho_{3}^{\alpha}\rho_{3}^{\beta}+\rho_{8}^{\alpha}\rho_{8}^{\beta}
 &=& \left(2 \sum_{l=1,2,3}
 |U_{\alpha l}|^2 
  |U_{\beta l}|^2 - \frac{2}{3} 
   \right)
  \nonumber \\
  &=& 
  \left(2 \left(
  \delta_{\alpha \beta}
  - 2 \sum_{k>l} 
  \mathfrak{Re} \left( U^{*}_{\alpha l}
   U_{\alpha k}
   U_{\beta l}
   U^{*}_{\beta k}
  \right) \right)- \frac{2}{3} 
   \right)\,.
   \label{rho3rho8}
\end{eqnarray}
Replacing the relations above, the neutrino oscillation probability turns out to be: 
\begin{equation}
\begin{split}
    P_{\nu_{\alpha}\rightarrow \nu_{\beta}}(t) &=
    \frac{1}{3} -\frac{1}{3}e^{-\Gamma_{1}t} + 
    \left(
    \delta_{\alpha \beta}
    - 2 \sum_{k>l} 
    \mathfrak{Re} \left( U^{*}_{\alpha l}
    U_{\alpha k}
    U_{\beta l}
    U^{*}_{\beta k}
    \right) \right) e^{-\Gamma_{1}t} \\ 
    &+ 2 
    \sum_{k>l} \mathfrak{Re} \left( U^{*}_{\alpha l}
    U_{\alpha k}
    U_{\beta l}
    U^{*}_{\beta k}
    \right) \cosh \left(\frac{\Omega_{kl}t}{2}\right)  e^{-\frac{1}{2}\Sigma_{12} t} \\
    &+ 
    2 
    \sum_{k>l}
    \left( -2 \, \Delta_{kl} \mathfrak{Im} \left( U^{*}_{\alpha l}
    U_{\alpha k}
    U_{\beta l}
    U^{*}_{\beta k}
    \right) 
    + \Delta\Gamma_{12} \mathfrak{Re} \left( U^{*}_{\alpha l}
    U_{\alpha k}
    U^{*}_{\beta l}
    U_{\beta k}  \right)
    \right)  \frac{\sinh{\left(\frac{\Omega_{kl}t}{2}\right)}}
    {\Omega_{kl}}
    e^{-\frac{1}{2}\Sigma_{12} t}\,.
\end{split}
\label{probsimp}
\end{equation}
It is straightforward to get the standard neutrino oscillation probability 
setting all the decoherence parameters to zero 
and using that in this case $\Omega_{kl} = 2 i \Delta_{kl}$. 

Isolating the quantum decoherence effects from the standard neutrino oscillation probability is useful for a better understanding of such kinds of effects. This becomes manageable since the approximation $\Omega_{kl} \approx 2 i \Delta_{kl}$ is 
valid for the energy range (few GeV) and size of decoherence effects we will study here. As a consequence, we can rearrange Eq.~(\ref{probsimp}) as follows: 
\begin{equation}
\begin{split}
    P_{\nu_{\alpha}\rightarrow \nu_{\beta}}(t) &=
    P^{osc}_{\nu_{\alpha}\rightarrow \nu_{\beta}}(t) -
   \frac{1}{3}\left(e^{-\Gamma_{1}t}-1 \right) + 
    \left(
    \delta_{\alpha \beta}
    - 2 \sum_{k>l} 
    \mathfrak{Re} \left( U^{*}_{\alpha l}
    U_{\alpha k}
    U_{\beta l}
    U^{*}_{\beta k}
    \right) \right) \left(e^{-\Gamma_{1}t}-1 \right) \\ 
    &+ 2 
    \sum_{k>l} \mathfrak{Re} \left( U^{*}_{\alpha l}
    U_{\alpha k}
    U_{\beta l}
    U^{*}_{\beta k}
    \right) \cos\left(\Delta_{kl}t\right) \left( e^{-\frac{1}{2}\Sigma_{12} t} -1 \right) \\
    &- 2 \, 
    \sum_{k>l}
    \mathfrak{Im} 
    \left( U^{*}_{\alpha l}
    U_{\alpha k}
    U_{\beta l}
    U^{*}_{\beta k}  
    \right)  
    \sin \left(\Delta_{kl}t\right)  
    \left( e^{-\frac{1}{2}\Sigma_{12} t} -1 \right)\\
    &+ 
   \Delta\Gamma_{12}
    \sum_{k>l}
    \mathfrak{Re} \left( U^{*}_{\alpha l}
    U_{\alpha k}
    U^{*}_{\beta l}
    U_{\beta k}  \right)
     \frac{\sin \left(\Delta_{kl}t\right)}{\Delta_{kl}}
    e^{-\frac{1}{2}\Sigma_{12} t}     
\end{split}
\label{probsimp3}
\end{equation}
where $ \mathfrak{Im}\left( U^{*}_{\alpha l} U_{\alpha k} U_{\beta l} U^{*}_{\beta k}\right)$ satisfies the well known identity
\[ J= - s_{\alpha\beta;kl} \, \; \mathfrak{Im} \left( U^{*}_{\alpha l} U_{\alpha k} U_{\beta l} U^{*}_{\beta k}\right)\,,\]
where $J$ is the Jarlskog invariant and $ s_{\alpha\beta;lk} =\pm 1$ (see Table 13.1 in Ref.~\cite{Giunti:2007ry} for details). It is worth mentioning that the accuracy of the above-mentioned neutrino oscillation probability is better than 0.1\%.

\subsubsection{Neutrino Oscillation Probability in Matter}
In this study, we will focus on two scenarios for $\delta_{\text{CP}}$: $180\degree$ and $270\degree$. Specifically, with our chosen $M_D^V$ and $\delta_{\text{CP}}= 180\degree$, the former remains invariant under rotation from the mass eigenstate basis in vacuum to the effective mass matter basis (diagonal basis). This implies that we can directly infer the matter neutrino oscillation probability from the vacuum neutrino oscillation probability by substituting the vacuum mixing angles and mass-squared differences with their respective values in matter~\cite{Carpio:2017nui}. For the case of $\delta_{\text{CP}}= 270\degree$, the decoherence matrix (defined in vacuum) is not invariant. Thus, we would have to make a nontrivial rotation to go to the (diagonal) effective matter basis, following the formalism described in Ref.~\cite{Carpio:2017nui}. However, given that this procedure can become cumbersome and does not particularly add anything different from the point of view of the phenomenology of the neutrino oscillation probability, we consider it unnecessary to display it here.

As follows, we apply the reasoning discussed above to derive one of the dominant probability channels in our analysis: $P_{\nu_{\mu}\rightarrow \nu_{e}}$ for $\delta_{\text{CP}}= 180\degree$. Therefore, the main goal is to evaluate the vacuum neutrino oscillation probability at $\delta_{\text{CP}}= 180\degree$ using Eq.~(\ref{probsimp3}). When $\delta_{\text{CP}}= 180\degree$, the following identities are satisfied:
 
\[ 
\mathfrak{Re} \left( U^{*}_{\mu l}
    U_{\mu k}
    U_{e l}
    U^{*}_{e k}
    \right) =\mathfrak{Re} \left( U^{*}_{\mu l}
    U_{\mu k}
    U^{*}_{e l}
    U_{e k}  \right)
    = \mathfrak{Re} \left( U_{\mu l}
    U_{\mu k}
    U_{e l}
    U_{e k} \right),
    \,
   \text{ and } 
   \mathfrak{Im} 
    \left( U^{*}_{\mu l}
 U_{\mu k} 
 U_{e l}
 U^{*}_{e k}\right)=0\,.
\]
Further simplification can be achieved by assuming that the terms proportional to either $\Delta\Gamma_{12} \sin^2{\theta_{13}}$ or $\Sigma_{12} \sin^2{\theta_{13}}$ are negligible in the neutrino oscillation probability. This allows us to write the approximate expressions:  
\[
    \mathfrak{Re} \left( U_{\mu 1} U_{\mu 2} U_{e 1} U_{e 2} \right) \approx \mathcal{D}\,,\,\, \,\,\,\,\,\,
    \mathfrak{Re} \left( U_{\mu 1} U_{\mu 3} U_{e 1} U_{e 3} \right) \approx \frac{1}{8} \tilde{J}\,, \,\, \,\,\,\,\,\,
    \mathfrak{Re} \left( U_{\mu 2} U_{\mu 3} U_{e 2} U_{e 3} \right) \approx -\frac{1}{8} \tilde{J}\,,
\]
where $\tilde{J} =\cos\theta_{13} \sin 2\theta_{12} \sin 2\theta_{13} \sin 2\theta_{23}$
and  ${\cal{D}} = \frac{1}{4} \sin 2\theta_{12} 
    \left (\cos 2\theta_{12} \sin 2\theta_{23} \sin \theta_{13} -
    \sin 2\theta_{12} \cos^2 \theta_{23} \right) $.
Equipped with all these ingredients, the vacuum version of $P_{\nu_{\mu}\rightarrow \nu_{e}}$ can be written as
\begin{equation}
\begin{split}
    P_{\nu_{\mu}\rightarrow \nu_{e}}(t) =
    P^{osc}_{\nu_{\mu}\rightarrow \nu_{e}}(t) 
    &-\left(\frac{1}{3}+ 2 D \right)\left(e^{-\Gamma_{1}t}-1 \right) \\ 
    &+ 
     \left(\frac{\tilde{J}}{4} \left( \cos \left(\Delta_{31} t\right)-\cos \left(\Delta_{32} t\right)\right)
    + 2 D \cos \left(\Delta_{21} t \right) \right)  \left( e^{-\frac{1}{2}\Sigma_{12} t} -1 \right)
   \\
    &+  
  \left.   \Delta\Gamma_{12}\left(\frac{\tilde{J}}{8}  \left( 
     \frac{\sin\left(\Delta_{31} t\right)}{\Delta_{31}} 
     - \frac{ \sin\left(\Delta_{32} t\right)}{\Delta_{32}} \right)   
     + D \frac{\sin \left(\Delta_{21}t\right)}{\Delta_{21}} \right)\right) 
    e^{-\frac{1}{2}\Sigma_{12} t}\,.
\end{split}
\label{probmue}
\end{equation}
Making the appropriate replacements of the mixing angles and squared mass differences (see Appendix~\ref{sec:AppendixB}), we have that $P_{\nu_{\mu}\rightarrow \nu_{e}}$ takes the form 
\begin{equation}
\begin{split}
    P_{\nu_{\mu}\rightarrow  \nu_{e}} & (L) =   
    P^{osc}_{\nu_{\mu}\rightarrow \nu_{e}}(L)
    -\left(\frac{1}{3}+ 2 D_m \right)\left(e^{-\Gamma_{1}L}-1 \right) \\ 
    &+ \left(\frac{\tilde{J_m}}{4} \left(\cos \left(\bar{\Delta}_{31}-2a L \right) -\cos \left(\bar{\Delta}_{31}-\bar{\Delta}_{21} \cos^2\theta_{12} \right)\right)
    + 2 D_m \cos \left(2a L \right) \right)  \left( e^{-\frac{1}{2}\Sigma_{12} L} -1 \right)
   \\
    &+  
    \Delta\Gamma_{12} \left(\frac{\tilde{J}_m}{8} 
    \left( 
     \frac{\sin \left(\bar{\Delta}_{31}-2a L \right)}{\left(\Delta_{31}-2a L \right)}
      - \frac{\sin \left(\bar{\Delta}_{31}-\bar{\Delta}_{21} \cos^2\theta_{12} \right)}{\left(\Delta_{31}-\Delta_{21} \cos^2\theta_{12} \right)}  \right)    
     + D_m  
    \frac {\sin\left(2a L\right)}{2a}
     \right) 
    e^{-\frac{1}{2}\Sigma_{12} L}\,, 
\end{split}
\label{probmuematter}
\end{equation}
where
\begin{eqnarray}
 \tilde{J}_m &=& - \cos\theta_{13} \sin 2\theta_{12} \sin 2\theta_{13} \sin 2\theta_{23} 
\left(\frac{\bar{\Delta}_{21}}{2aL} \right)
    \left(\frac{\bar{\Delta}_{31}}{\bar{\Delta}_{31}-2aL} \right)  \nonumber \\
   &=& -\tilde{J} \left(\frac{\bar{\Delta}_{21}}{2aL} \right)
    \left(\frac{\bar{\Delta}_{31}}{\bar{\Delta}_{31}-2aL} \right) 
\end{eqnarray}
and 
\begin{eqnarray}
    {D_m} &=&
    -\frac{\sin 2\theta_{12}}{4} \left(
    \left(\frac{\bar{\Delta}_{21}}{2aL} \right)^2 \sin 2\theta_{12} \cos^2\theta_{23} + 
\left(\frac{\bar{\Delta}_{21}}{2aL} \right) \left(\frac{\bar{\Delta}_{31}}{\bar{\Delta}_{31}-2aL} \right) \sin \theta_{13}
  \sin 2\theta_{23} \right)
\end{eqnarray}
with ${\bar{\Delta}_{ij}}=\Delta_{ij}L$ and  $a=\frac{G_F N_e}{\sqrt{2}}$, where $G_F$ is the Fermi constant and $N_e$ is the electron number density. It should be noted that, according to Eq.~(\ref{probmuematter}), most of the terms that involve decoherence parameters are entangled with matter effects, which makes them CP-odd terms. This extra CP-odd contribution will be reflected in our results.  

In the analysis we also consider other values of $\delta_{\text{CP}}$. As explained above, the decoherence matrix (defined in vacuum) is not invariant for this value of the CP phase, and the calculation would become cumbersome and is hence avoided since it does not add anything to the phenomenological discussion of this section. It should be noted, however, that the sensitivity results presented below use numerical (exact) calculations of the neutrino oscillation probabilities for any value of $\delta_{\text{CP}}$.

\begin{figure}[t]
  \centering
    \includegraphics[width=0.95\linewidth]{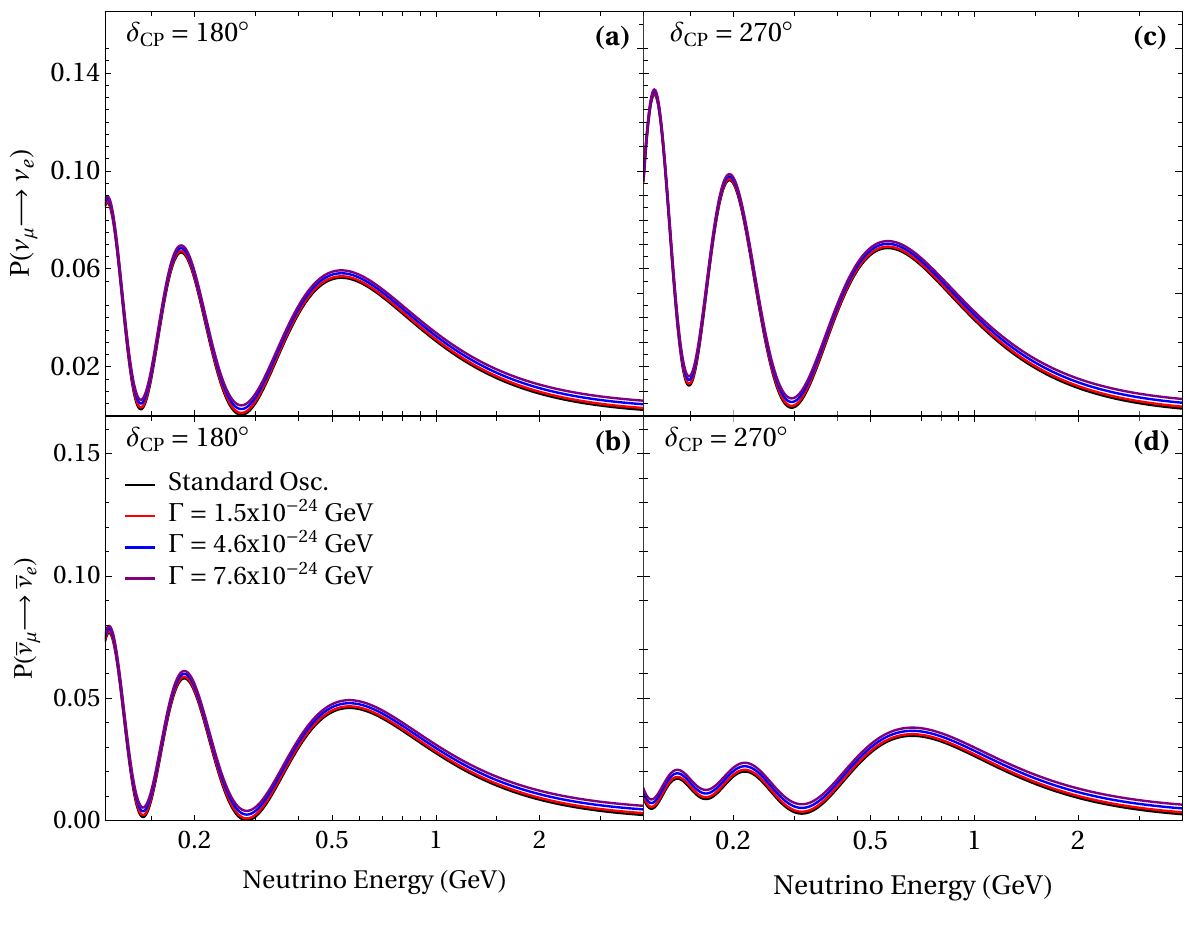}
    \caption{The neutrino oscillation probability at T2HK. Panels (a) and (b) are neutrino and anti-neutrino probabilities for $\delta_{\text{CP}} = 180\degree$, respectively, and panels (c) and (d) are neutrino and anti-neutrino probabilities for $\delta_{\text{CP}} = 270\degree$, respectively.}
    \label{fig:T2HK-Prob}
\end{figure}

\begin{figure}[t]
  \centering
    \includegraphics[width=0.95\linewidth]{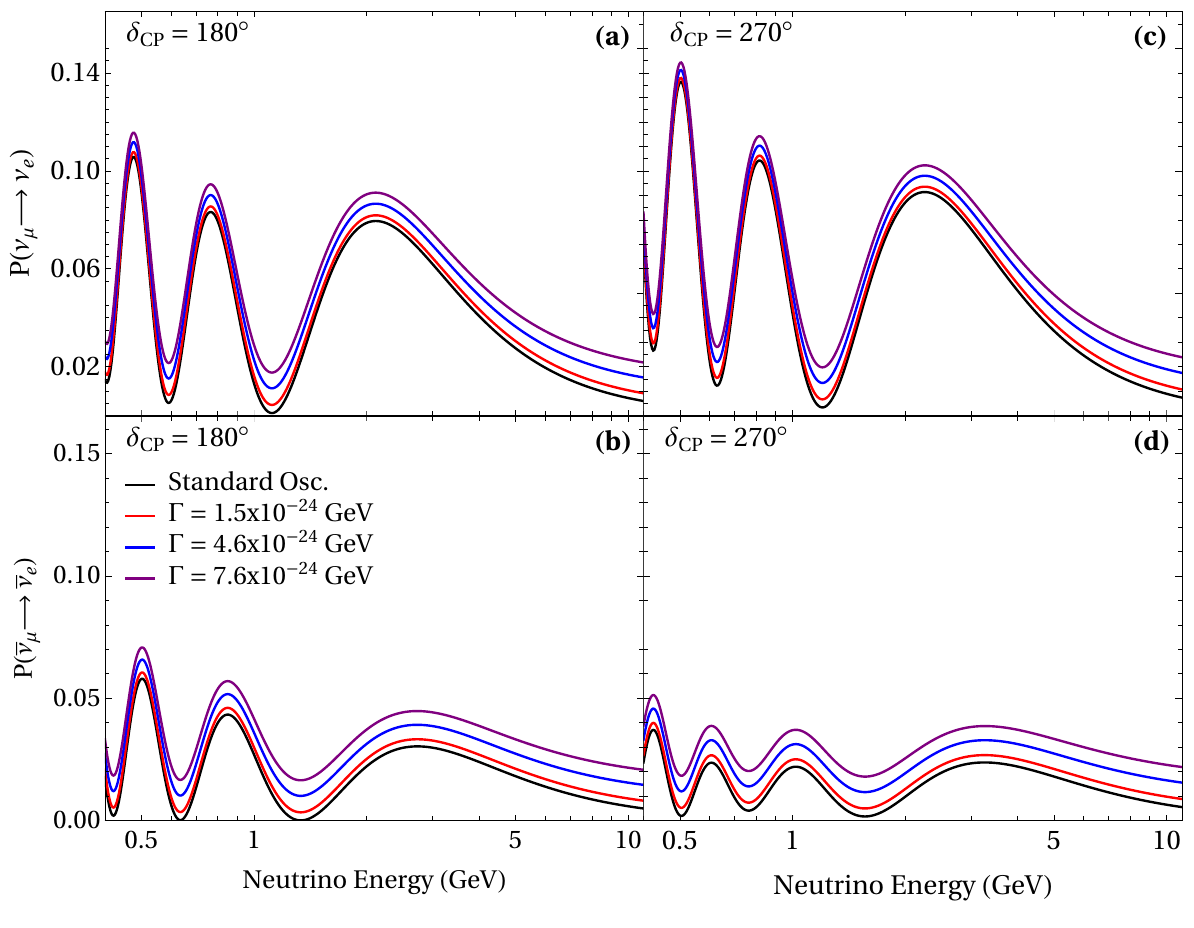}
    \caption{The neutrino oscillation probability at DUNE. Panels (a) and (b) are neutrino and anti-neutrino probabilities for $\delta_{\text{CP}} = 180\degree$ respectively, and panels (c) and (d) are neutrino and anti-neutrino probabilities for $\delta_{\text{CP}} = 270\degree$, respectively.}
    \label{fig:DUNE-Prob}
\end{figure}

\begin{figure}[t]
  \centering
    \includegraphics[width=0.95\linewidth]{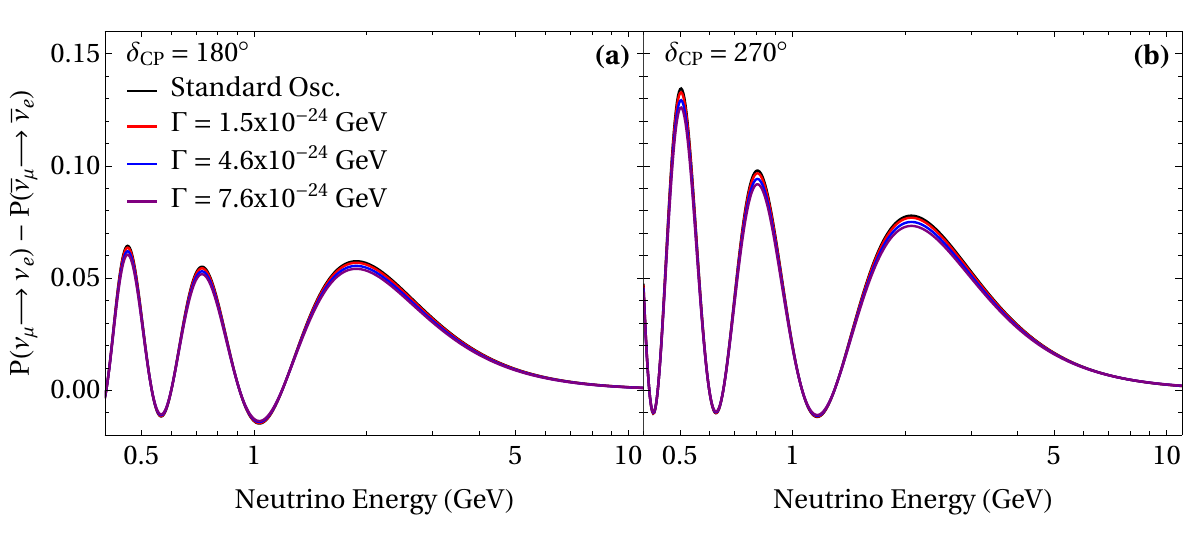}
    \caption{DUNE Asymmetry. Panels (a) and (b) are the asymmetry for $\delta_{\text{CP}} = 180\degree$ and $\delta_{\text{CP}} = 270\degree$, respectively.}
    \label{fig:DUNE-Asymm}
\end{figure}

\begin{table}[t]
    \centering
    \begin{tabular}{|c|c|}
    \hline
         $\Delta m_{21}^2$ & $7.50\times10^{-5}$~eV$^2$  \\
         $\Delta m_{31}^2$ & $2.55\times10^{-3}$~eV$^2$  \\
         \hline
         $\sin^2\theta_{12}$ & $0.32$ \\              
         $\sin^2\theta_{13}$ & $0.022$ \\
         $\sin^2\theta_{23}$ & $0.57$ \\
    \hline
         $\delta_{\text{CP}}$ & $270\degree$, $180\degree$ \\               
    \hline
    \end{tabular}
    \caption{\label{tab:osc_par} Standard neutrino oscillation parameters used to generate the fake data sets in this paper.}
\end{table}

In Figs.~\ref{fig:T2HK-Prob} and~\ref{fig:DUNE-Prob} we display the neutrino and antineutrino oscillation probabilities for $\delta_{\text{CP}} = 180\degree$ and $\delta_{\text{CP}} = 270\degree$, and for source-detector distances of $L=295$~km, and $L=1284.9$~km, and matter densities $\rho = 2.60$~g/cm$^3$ and $\rho = 2.848$~g/cm$^3$, compatible with the ones corresponding to T2HK  and DUNE, respectively.
The remaining standard neutrino oscillation parameters are those summarized in Tab.~\ref{tab:osc_par}, taken from Ref.~\cite{deSalas:2020pgw}. We fix $\Gamma_2= \frac{5}{3}\Gamma_1$, corresponding to the upper limit from the inequality discussed above. In each panel we present four cases $\Gamma:=\Gamma_1=1.5 \times 10^{-24}$, $4.6 \times 10^{-24}$, $7.6 \times 10^{-24}$ GeV and the standard oscillation case. As expected, for the T2HK experiment, the quantum decoherence effects are negligible. This happens because $\Gamma L$ is very small for the T2HK source-detector distance, $L = 295$~km. The T2HK experiment will be helpful for the determination of the oscillation parameter without being distorted by quantum decoherence effects. This distortion happens when the fitting is made under a theoretical hypothesis based on pure standard neutrino oscillation probability, a situation considered in this paper.  

In contrast, the quantum decoherence effects are visible for the neutrino oscillation probability for the source-detector distance at DUNE. For instance, the almost energy-independent increase of the neutrino/antineutrino oscillation probabilities is a pattern already noticed in Ref.~\cite{Carpio:2017nui}. This behavior comes from the extra terms involving the decoherence parameters, a contribution mainly dominated by the constant term $-\frac{1}{3}\left(e^{-\Gamma_{1}L}-1 \right)$, which can be identified in Eq.~(\ref{probmuematter}) for the case of $\delta_{\text{CP}}=180\degree$. Another noticeable feature is how the neutrino and the antineutrino oscillation probabilities are becoming more similar as $\Gamma$ grows. This trend is more clearly visible in Fig.~\ref{fig:DUNE-Asymm}, which shows $\Delta P = P(\nu_\mu \rightarrow \nu_e) -P(\bar{\nu}_\mu \rightarrow \bar{\nu}_e)$, where the tendency of $\Delta P $ is to decrease when $\Gamma$ increases. The $\Delta P$ differences 
between the non-zero $\Gamma$ cases and the standard oscillation case are small for both $\delta_{\text{CP}}=180\degree$ and $\delta_{\text{CP}}=270\degree$.
This observation suggests that an eventual fitted value of $\delta_{\text{CP}}$, considered solely standard neutrino oscillation as a theoretical hypothesis, would not be far from its true value even if decoherence effects are present in nature. The latter would be expected if, in the data analysis, most of the standard neutrino oscillation parameters were fixed in the fit. However, our analysis has a wider perspective, contemplating that $\delta_{\text{CP}}$, $\theta_{13}$, $\theta_{23}$, and $\Delta m^2_{31}$ can vary. In this way, and again taking neutrino standard oscillation as a theoretical hypothesis, the possibility of finding bigger distortions between the fitted values of $\delta_{\text{CP}}$ and its true value is open. This argumentation is also valid for the other fitted standard neutrino oscillation parameters.

\section{Simulation details}
\label{sec:sim}

In the following sections, we will employ the theoretical discussion of the last section in experimental context and discuss how decoherence effects could affect precision measurements at future facilities. In our analyses, we consider the T2HK~\cite{Hyper-KamiokandeProto:2015xww} and DUNE experiments~\cite{DUNE:2021cuw,DUNE:2020lwj,DUNE:2020ypp,DUNE:2020mra,DUNE:2020txw}. In the following subsections, we briefly discuss some details on the simulation for both experiments. We use GLoBES~\cite{Huber:2004ka,Huber:2007ji} for the numerical analyses performed in this paper. 

\subsection{T2HK}
\label{sec:sim_t2hk}

The currently operating Super-K detector will be in a few years replaced by a larger version containing $\sim250$~kt of water. This larger detector is going to measure neutrinos from the Sun, from the atmosphere, and also from a neutrino beam generated at J-Parc as is currently the case in T2K. The updated version of the experiment is called T2HK (Tokai-to-Hyper-Kamiokande).
The necessary information for the simulation of T2HK is extracted from Refs.~\cite{NOW22,NuFACT23}.
The computational simulation was performed using the GLoBES package \cite{Huber:2004ka,Huber:2007ji} and is based on the predefined AEDL file from Ref.~\cite{GLoBEST2HK}. We consider only quasi-elastic interactions due to the energy range under consideration. We have replaced the flux files from Ref.~\cite{GLoBEST2HK} with those published by the T2K collaboration in Ref.~\cite{T2K:2023smv}. We use the spectral form of these fluxes and change the normalization to the one expected at Hyper-K.
We assume 7.5 years of data taking in antineutrino mode and 2.5 years in neutrino mode, and include both disappearance and appearance channels in our analyses. In order to obtain a reasonable approximation for the resolution function (not provided by the collaboration and needed in GLoBES), we calculate the sensitivity to CP violation and compare it with the corresponding plot in Ref.~\cite{NOW22} for different choices of systematic uncertainties. We find that we can reproduce the curves using $\sigma_E = aE + b\sqrt{E}$ with $a=0.031$ and $b=0.0822$ for all channels. Background components, relevant for our analysis, are extracted from Ref.~\cite{NOW22}. Systematic uncertainties in the signal and background are taken into account in our analysis and are taken from Ref.~\cite{NOW22}. The same uncertainty was used for the signal and background components in each channel, with values of $2.56\%$ for $\nu_e$ appearance, $2.53\%$ for $\bar{\nu}_e$ appearance, $1.89\%$ for $\nu_\mu$ disappearance, and $1.74\%$ for $\bar{\nu}_\mu$ disappearance.

\subsection{DUNE}
\label{sec:sim_dune}

DUNE~\cite{DUNE:2020lwj,DUNE:2020ypp,DUNE:2020mra,DUNE:2020txw}~ is an experiment under construction, and it will be the successor experiment to NOvA. It will consist of two detectors, which are going to be exposed to a megawatt-scale neutrino beam produced at
Fermilab, composed of (nearly) only muon neutrinos or antineutrinos. The
near and far detectors will be placed $\mathcal{O}(100)$~m and 1284.9~km away from the
source of the beam. The far detector will be divided into four
modules, each using 10~kton of argon as detection material.
To simulate the neutrino signal at DUNE we use the configuration file for GLoBES provided by the DUNE collaboration \cite{DUNE:2021cuw}, which assumes 6.5 years of running time in neutrino (FHC) and antineutrino (RHC) mode, a far detector with 40 kt fiducial mass of liquid argon, and a beam from collisions of 120 GeV protons and 1.2 MW beam power (i.e. 624 kt-MW-years of exposure). Within these files smearing matrices are included for a proper treatment of energy resolution.
We include disappearance and appearance channels, simulating both signals and backgrounds as suggested in Ref.~\cite{DUNE:2021cuw}. The simulated backgrounds include contamination of antineutrinos (neutrinos) in the neutrino (antineutrino) mode, and also misinterpretation of flavors. 
Regarding systematic uncertainties, we include nuisance parameters in the form of normalizations of the signal and background components in each channel, whose uncertainties range between 2\% and 20\% depending on the signal or background component.

\subsection{Statistical analysis}
\label{sec:stat_an}

When performing the statistical analysis, we need to include several sources of systematic uncertainties related to the signal and background predictions. In addition, correlations among the neutrino oscillation parameters must be included.
The $\chi^2$ function for the analysis of T2HK and DUNE pseudo-data is given by
\begin{equation}
\label{eq:T2KNOvAchi2}
 \chi^2_{\mathrm{T/D}}(\vec{p})=\min_{\vec{\alpha}}
 \sum_\text{channels}2\sum_i \left[ N_{\text{exp},i}(\vec{p},\vec{\alpha})- N_{\text{dat},i} +
 N_{\text{dat},i} \log \left(\frac{N_{\text{dat},i}}{N_{\text{exp},i}(\vec{p},\vec{\alpha})}\right)\right] 
 + \sum_i \left(\frac{\alpha_i}{\sigma_i}\right)^2 + \chi^2_{\text{solar}}\,,
\end{equation}
where $N_{\text{dat},i}$ denote the mock data per energy bin $i$, while $N_{\text{exp},i}(\vec{p},\vec{\alpha})$ indicate the expected event numbers for a given set of oscillation parameters $\vec{p}$ and systematic uncertainties $\vec{\alpha}$.
The first sum is taken over the different oscillation channels: $\nu_{\mu}\to\nu_{\mu}$, $\overline{\nu}_{\mu}\to\overline{\nu}_{\mu}$, $\nu_{\mu}\to\nu_e$ and $\overline{\nu}_{\mu}\to\overline{\nu}_e$. 
The second term contains penalty factors for all the systematic uncertainties $\alpha_k$, with expectation value $\mu_k = 0$ and standard deviation $\sigma_k$. 
The last term contains a penalty for the solar parameters: $\theta_{12}$ and $\Delta m_{21}^2$. We do not include a prior on $\theta_{13}$ in our analyses, since decoherence effects could affect the determination of this parameter, but we will compare our results with current measurements from reactor experiments.
We will also perform a combined analysis of T2HK and DUNE. In this case, we must first sum the $\chi^2$ contributions from each experiment and later perform the marginalization.

\begin{figure}[t!]
  \centering
    \includegraphics[width=0.95\linewidth]{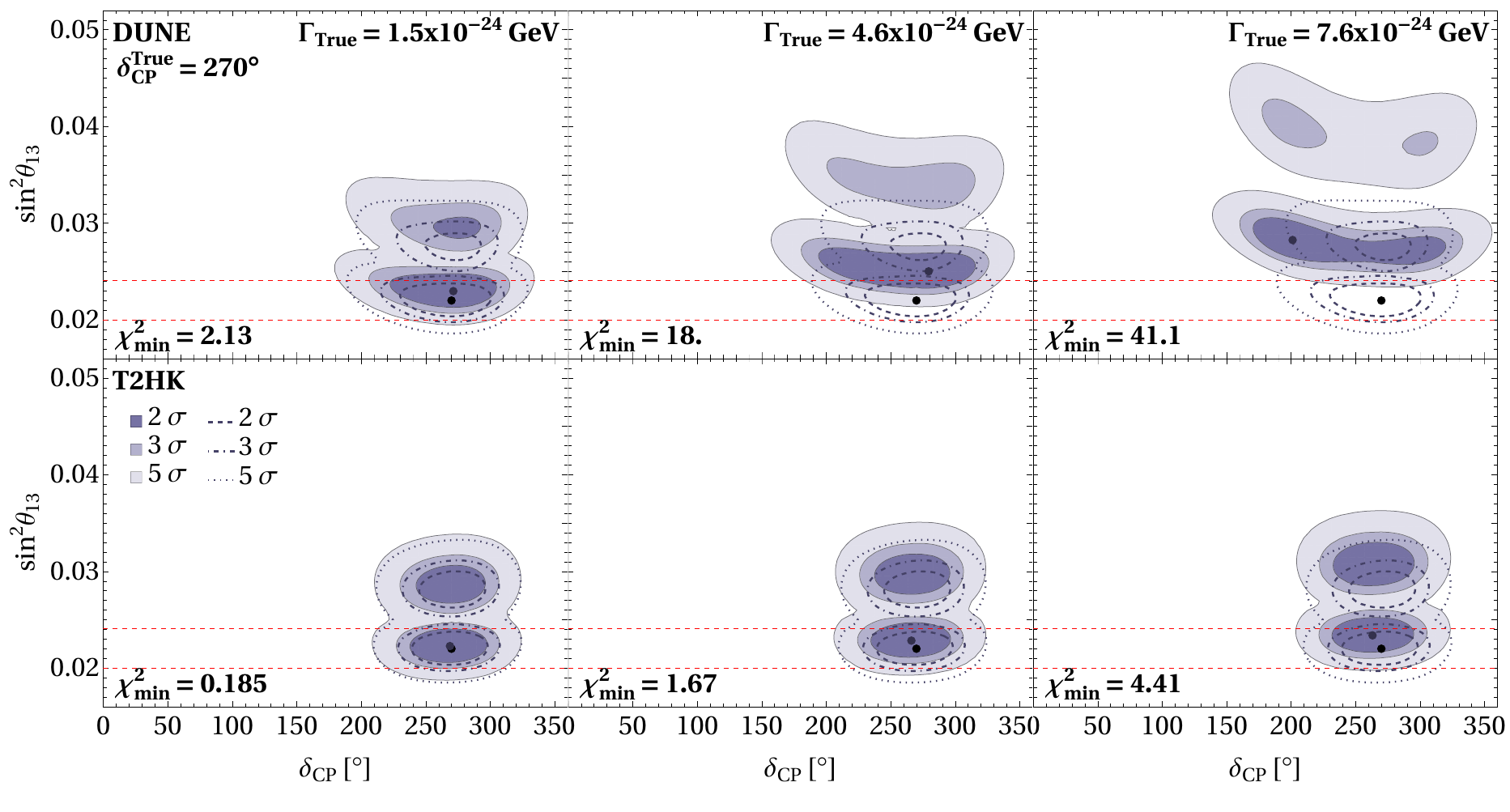}
    \caption{Expected contours of $\delta_{\text{CP}}$ vs. $\sin^2 \theta_{13}$ with $\delta_{\text{CP}}^{\text{True}}=270\degree$ for several choices of $\Gamma_\text{true}$. The upper row is for DUNE and the lower one for T2HK. The black dot indicates the true value assumed in the generation of mock data, and the gray one indicates the obtained best fit value. The horizontal red dashed lines represent the upper and lower $3\sigma$ limits of $\sin^2 \theta_{13}$ \cite{deSalas:2020pgw}. The filled regions correspond to when $\Gamma_{\text{true}} \neq 0$, while the 
    dashed/dot-dashed/dotted contour lines are the results of the standard sensitivities of the experiments shown for comparison.}
    \label{fig:DUNEyT2HK-dcp_th13-90}
\end{figure}

\section{Results and discussion}
\label{sec:res}

In this section we discuss the results of our simulations of DUNE and T2HK and of the combination of the two. We want to see how much of an effect neutrino quantum decoherence would have on the determination of standard oscillation parameters. In order to do this, we generate mock data sets with decoherence in the data, i.e. $\Gamma_\text{True} > 0$, and try to reconstruct these data using only standard parameters. When generating the mock data sets, we fix the standard neutrino oscillation parameters to those shown in Tab.~\ref{tab:osc_par}, taken from Ref.~\cite{deSalas:2020pgw}, except for $\delta_\text{CP}$, for which we assume $\delta_\text{CP}=270\degree$ and $\delta_\text{CP}=180\degree$, which are very close to the best fit parameters obtained by T2K~\cite{T2K:2023mcm} and NOvA~\cite{NOvA:2021nfi,NOvA:2023iam}, respectively.

In Fig.~\ref{fig:DUNEyT2HK-dcp_th13-90} we show the results in the $\delta_\text{CP}-\sin^2\theta_{13}$ plane at 2, 3 and 5$\sigma$ confidence level (C.L.) assuming $\delta_\text{CP}^\text{True}=270\degree$ and three different values for $\Gamma_\text{True}$. 
The upper (lower) row corresponds to our analysis of DUNE (T2HK). The dashed/dot-dashed/dotted contour lines correspond to the standard sensitivities ($\Gamma_\text{True} = 0$) of the experiments and are shown for comparison. 
The filled regions are obtained for the values of $\Gamma_\text{True}$ indicated in each panel. The minimal value of $\chi^2$ obtained in each analysis is also shown in each panel. 
Note that the regions of DUNE are more deformed (with respect to the standard analysis) than the ones of T2HK. This is due to the fact that the decoherence effect under consideration has a larger effect in DUNE than in T2HK due to the larger baseline. 
One can see how the region deforms and in the case of the right panel interestingly we find that even though the data was generated with maximal CP violation, the best fit is found very close to CP conservation.
It should be noted that also the allowed values of $\theta_{13}$ are effected. As can be seen in the right panel the determination of $\theta_{13}$ from accelerator data would be in tension with data from reactor experiments, indicated by the red dashed lines\footnote{Note, however, that these lines correspond to the standard analysis, and that the inclusion of decoherence effects might also effect the measurement of $\theta_{13}$ in reactor experiments.}. Therefore, if future long-baseline experiments measured a large value of $\theta_{13}$ the reason might be due to neutrino quantum decoherence.

\begin{figure}[t!]
  \centering
    \includegraphics[width=0.95\linewidth]{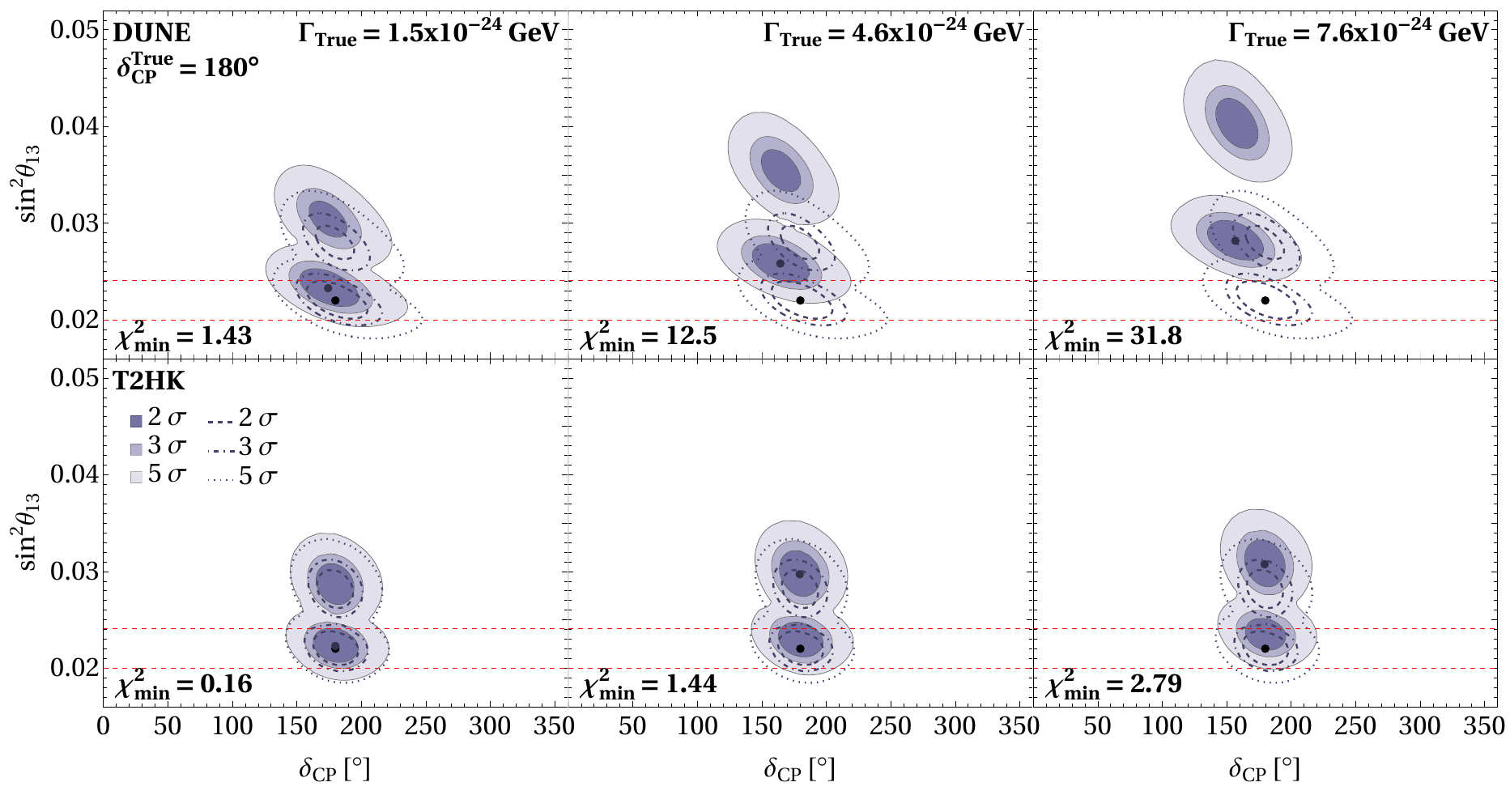}
    \caption{As Fig.~\ref{fig:DUNEyT2HK-dcp_th13-90} but for $\delta_{\text{CP}}^{\text{True}}=180\degree$.}
    \label{fig:DUNEyT2HK-dcp_th13-180}
\end{figure}

In Fig.~\ref{fig:DUNEyT2HK-dcp_th13-180} we show the results assuming $\delta_{\text{CP}}^{\text{True}}=180\degree$. 
The minimal values of $\chi^2$ obtained in these analyses are slightly smaller than the ones for $\delta_{\text{CP}}^{\text{True}}=270\degree$. The shift in the best fit value of $\delta_\text{CP}$ is also smaller than in the previous case. 
This happens due to the fact that a mock data set generated with $\delta_{\text{CP}}^{\text{True}}=270\degree$ induces maximally opposite shifts in $\nu_e$ and $\overline{\nu}_e$ events, while a data set generated with $\delta_{\text{CP}}^{\text{True}}=180\degree$ induces non-significant differences in the two spectra (from neutrino oscillations). 
A finite $\Gamma_\text{true}$ affects both $\nu_e$ and $\overline{\nu}_e$ spectra identically. Therefore a "deformed" spectrum with $\delta_{\text{CP}}^{\text{True}}=270\degree$ and finite $\Gamma_\text{true}$ is more difficult to reconstruct using only standard neutrino oscillation parameters than a spectrum with $\delta_{\text{CP}}^{\text{True}}=180\degree$ and finite $\Gamma_\text{true}$. 
However, we observe similar behavior in both cases for $\theta_{13}$. As previously the effect for the $\Gamma_\text{True}$ under consideration is rather small for T2HK.

\begin{figure}[t!]
  \centering
    \includegraphics[width=0.95\linewidth]{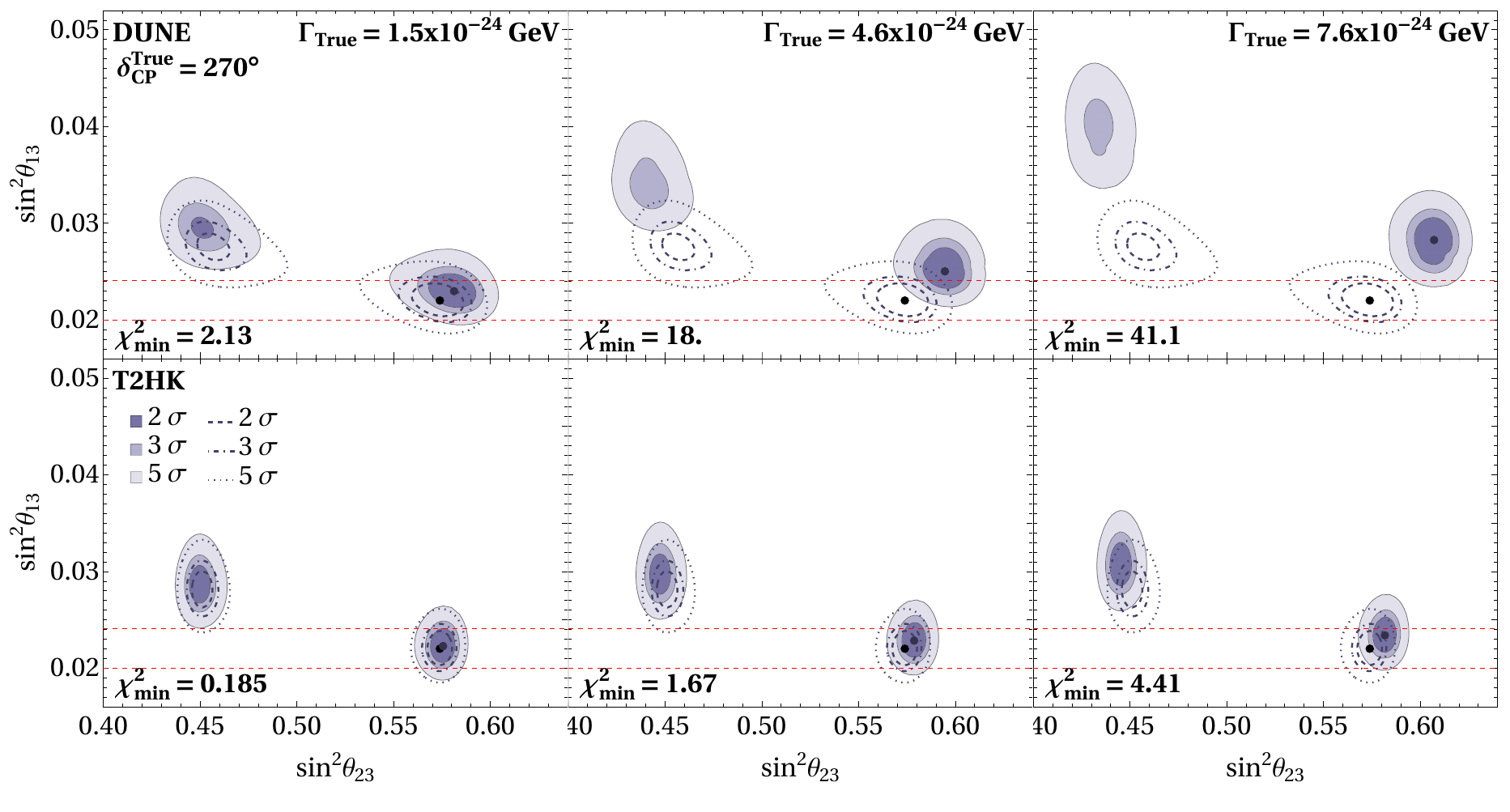}
    \caption{Expected contours of $\sin^2 \theta_{23}$ vs. $\sin^2 \theta_{13}$ with $\delta_{\text{CP}}^{\text{True}}=270\degree$. The upper row is for DUNE and the lower one  for T2HK.}
    \label{fig:DUNEyT2HK-th23_th13-90}
\end{figure}

\begin{figure}[t!]
  \centering
    \includegraphics[width=0.95\linewidth]{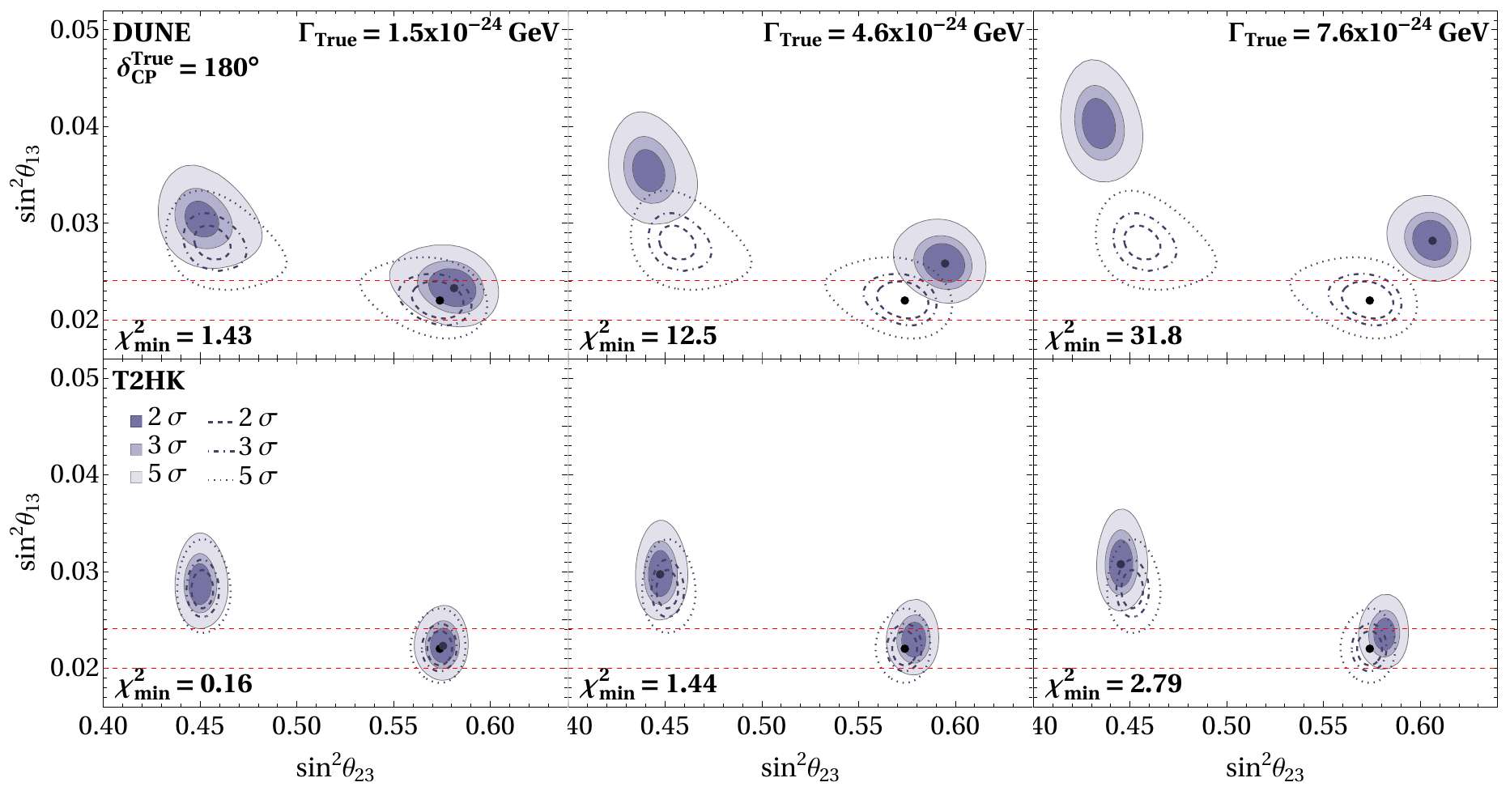}
    \caption{As Fig.~\ref{fig:DUNEyT2HK-th23_th13-90} but using $\delta_{\text{CP}}^{\text{True}}=180\degree$.}
    \label{fig:DUNEyT2HK-th23_th13-180}
\end{figure}

In Figs.~\ref{fig:DUNEyT2HK-th23_th13-90} and~\ref{fig:DUNEyT2HK-th23_th13-180} we show the equivalent to Figs.~\ref{fig:DUNEyT2HK-dcp_th13-90} and~\ref{fig:DUNEyT2HK-dcp_th13-180}, but in the $\sin^2\theta_{23}-\sin^2\theta_{13}$ plane for $\delta_{\text{CP}}^{\text{True}}=270\degree$ (Fig.~\ref{fig:DUNEyT2HK-th23_th13-90}) and $\delta_{\text{CP}}^{\text{True}}=180\degree$ (Fig.~\ref{fig:DUNEyT2HK-th23_th13-180}). 
The behavior is very similar for both cases. As mentioned, the preferred value for $\sin^2\theta_{13}$ is shifted to large values. At the same time the preferred values for $\sin^2\theta_{23}$ are shifted away from maximal mixing, $\sin^2\theta_{23} = 0.5$. The island in the lower octant of $\theta_{23}$ is shifted towards smaller values, while the island in the upper octant is shifted towards larger values. This happens because if decoherence is included in the data, the spectra can be reconstructed with smaller values of $\sin^22\theta_{23}$ (note the 2) than in the standard case, since decoherence leads to a damping in the neutrino oscillation probability.

From Figs.~\ref{fig:DUNEyT2HK-dcp_th13-90}--~\ref{fig:DUNEyT2HK-th23_th13-180} we see that if nature is such that neutrinos decohere in accelerator experiments, the measurement of standard neutrino oscillation parameters would be mostly unaffected at T2HK, while large distortions could be expected at DUNE. However, in the case of DUNE the data would not be fit as well anymore as in the case of T2HK as can be seen from the larger values of $\chi^2_\text{min}$ obtained in the analyses of DUNE. This effect can be further enhanced by performing a combined analysis of T2HK and DUNE.

\begin{figure}[t!]
  \centering
    \includegraphics[width=0.95\linewidth]{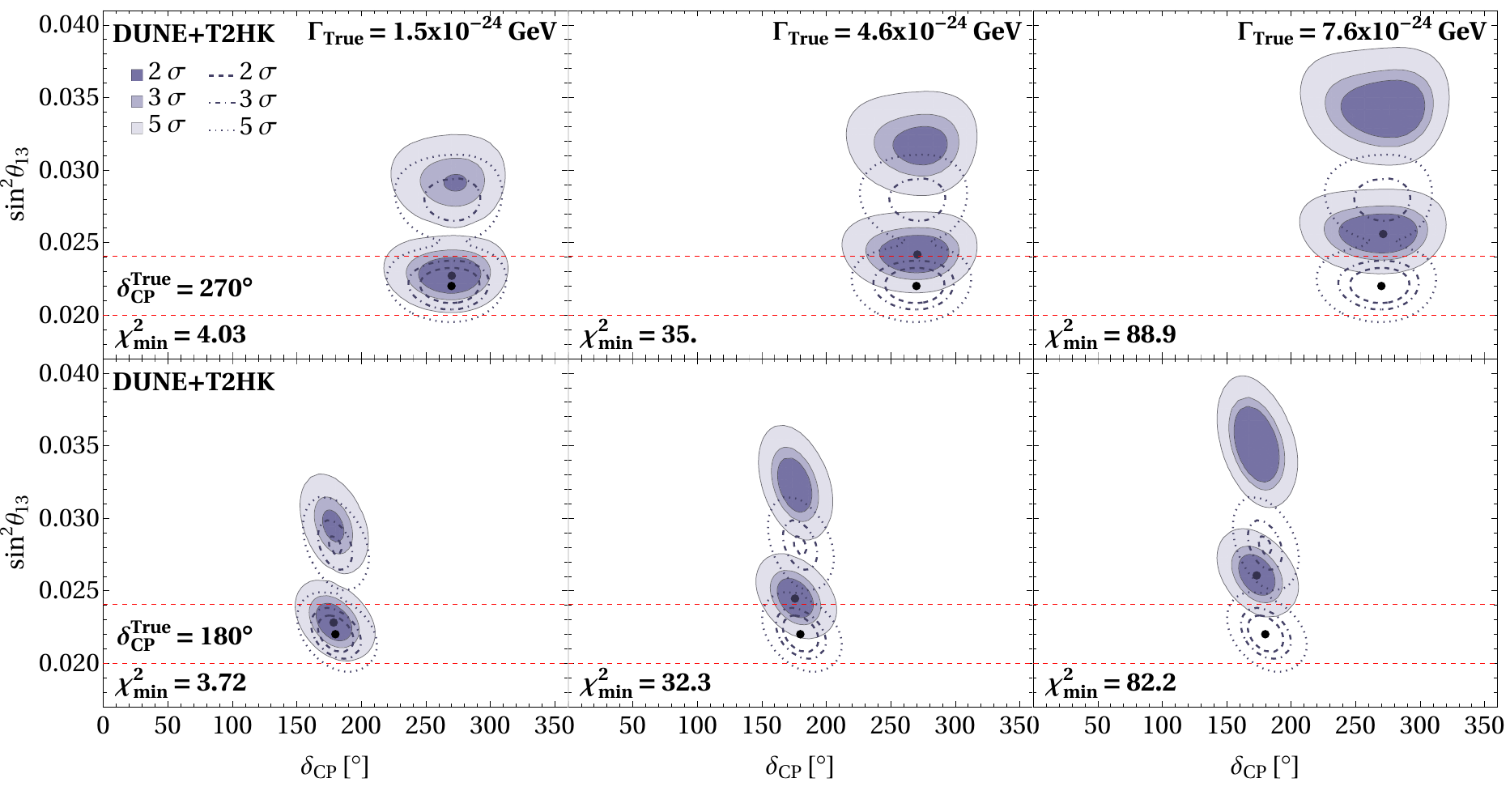}
    \caption{Expected contours of $\delta_{\text{CP}}$ vs. $\sin^2 \theta_{13}$ with $\delta_{\text{CP}}^{\text{True}}=270\degree$ for the upper row plots and $\delta_{\text{CP}}^{\text{True}}=180\degree$ for the lower row plots.}
    \label{fig:DUNEplusT2HK-dcp_th13}

\end{figure}

\begin{figure}[t!]
  \centering
    \includegraphics[width=0.95\linewidth]{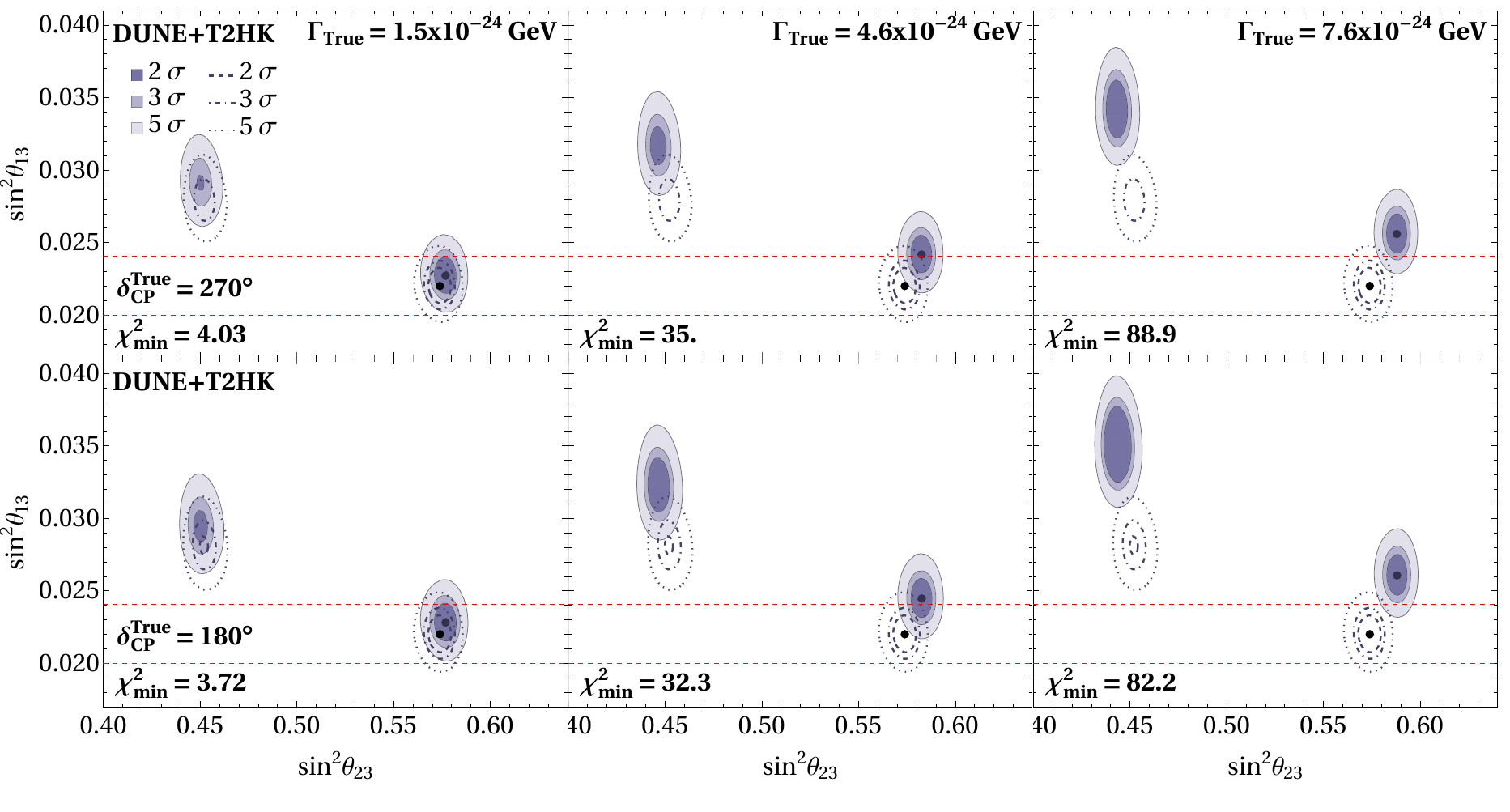}
    \caption{Expected contours of $\sin^2 \theta_{23}$ vs. $\sin^2 \theta_{13}$ with $\delta_{\text{CP}}^{\text{True}}=270\degree$ for the upper row plots and $\delta_{\text{CP}}^{\text{True}}=180\degree$ for the lower row plots.}
    \label{fig:DUNEplusT2HK-th23_th13}

\end{figure}

The results of this combined analysis are shown in Fig.~\ref{fig:DUNEplusT2HK-dcp_th13} in the $\delta_\text{CP}-\sin^2\theta_{13}$ plane and in Fig.~\ref{fig:DUNEplusT2HK-th23_th13} in the $\sin^2\theta_{23}-\sin^2\theta_{13}$ plane. From Fig.~\ref{fig:DUNEplusT2HK-dcp_th13} is becomes immediately clear that when performing a combined analysis of the experiments a robust measurement of the CP phase becomes possible. As can be seen in all panels the measurement of $\delta_\text{CP}$ is unaffected by the presence of decoherence effects in the data. 

The preferred value of $\theta_{13}$ is still shifted towards larger values. However, note that the $\chi^2_\text{min}$ obtained in the analysis are now much larger than before. Therefore, if such a result were observed once real data taking begins, it could hint towards decoherence effects in neutrino oscillations.
As can be seen in Fig.~\ref{fig:DUNEplusT2HK-th23_th13} the same argument applies to the measurement of $\sin^2\theta_{23}$.
We can therefore conclude that if a combined analysis is performed, a robust measurement of the standard neutrino oscillation parameters is possible. In addition, if such a situation appeared in real data, due to the goodness of the fit, one might start to explore new physics models in order to obtain a better fit to the data. 

\section{Conclusions}
\label{sec:conc}
In this paper we have discussed neutrino quantum decoherence effects in future long-baseline accelerator experiments. We discussed these in an analytic way in Sec.~\ref{sec:theo} and on experimental level in Secs.~\ref{sec:sim} and~\ref{sec:res}. In 
Sec.~\ref{sec:theo} we derived approximated analytical formulas for the neutrino oscillation probabilities in matter seeing the nearly energy-independent rise behavior in the probability, relative to the standard oscillation case. From the neutrino oscillation probability, the CP odd nature from the terms involving the quantum decoherence parameters and matter effects becomes evident. 

Next, we discussed how the measurement of standard oscillation parameters could be affected by the presence of decoherence effects. To quantify this, we generated mock data sets including decoherence and analyzed these data assuming standard 3-neutrino oscillations. For the decoherence parameters under consideration we have found that the measurement at DUNE is strongly affected, while the measurement at T2HK is not. This happens due to the much larger baseline at DUNE. 
As discussed in Sec.~\ref{sec:res} the inclusion of decoherence shifts the best fit of $\delta_\text{CP}$ and also worsens its determination. In addition, the inclusion of decoherence effects drives $\sin^2\theta_{13}$ towards larger values. Similarly, $\sin^2\theta_{23}$ values further away from maximal mixing become preferred. In the case of T2HK a similar effect can be observed, although in a much weaker way.

We have also performed a combined analysis of T2HK and DUNE. We have shown that this is a very powerful combination, because in the analyses with large decoherence effects very large values of $\chi^2_\text{min}$ are found. One can therefore argue that the combined analysis could provide a robust measurement of standard oscillation parameters, while also placing strong bounds on decoherence effects.

\section{Acknowledgements}

\noindent
This work is supported by the Spanish grants  CIPROM/2021/054 (Generalitat Valenciana) and PID2020-113775GB-I00 (AEI/10.13039/501100011033), by the European ITN project HIDDeN (H2020-MSCA-ITN-2019/860881-HIDDeN), 
Marie Skłodowska-Curie Staff Exchange grant  ASYMMETRY (HORIZON-MSCA-2021-SE-01/101086085-ASYMMETRY) and by
{\it{Dirección
de Fomento de la Investigación}} at Pontificia Universidad
Católica del Perú, through Grants No. DFI-2021-0758 and CONCYTEC through Grant
No.60-2015-FONDECYT. A.M. Gago want to thank José Carpio for useful discussions. A. Catalayud and  A.M. Gago want to express their gratitude to the Departament de Física Teòrica, Universitat de València for their warm hospitality during the development of this work.

\appendix
\section{$\rho_\mu^\alpha$ definitions}
\label{sec:AppendixA} 
The coefficients $\rho_{\mu}^\alpha$
are  
written as
\begin{eqnarray}
& & \rho_{0}^{\alpha}=\sqrt{2/3} \nonumber \\
& & \rho_{1}^{\alpha}=2\;\Re(U_{\alpha 1}^{\ast}U_{\alpha 2}) \nonumber \\
& & \rho_{2}^{\alpha}=-2\;\Im(U_{\alpha 1}^{\ast}U_{\alpha 2})\nonumber \\
& & \rho_{3}^{\alpha}=|U_{\alpha 1}|^{2}-|U_{\alpha 2}|^{2} \nonumber \\
& & \rho_{4}^{\alpha}=2\;\Re(U_{\alpha 1}^{\ast}U_{\alpha 3}) \\
& & \rho_{5}^{\alpha}=-2\;\Im(U_{\alpha 1}^{\ast}U_{\alpha 3}) \nonumber \\
& & \rho_{6}^{\alpha}=2\;\Re(U_{\alpha 2}^{\ast}U_{\alpha 3}) \nonumber \\
& & \rho_{7}^{\alpha}=-2\;\Im(U_{\alpha 2}^{\ast}U_{\alpha 3}) \nonumber \\
& & \rho_{8}^{\alpha}=\frac{1}{\sqrt{3}}(|U_{\alpha 1}|^{2}+|U_{\alpha 2}|^{2}-2|U_{\alpha 3}|^{2}) \nonumber
\end{eqnarray}

\section{Effective mixing angles and mass-squared differences in matter}
\label{sec:AppendixB} 
We are following \cite{Nunokawa:2007qh} and write the mixing angles and squared mass differences up to first order in $\alpha,\theta_{13}$ and $\alpha = \Delta m^2_{21}/\Delta m^2_{31}$ as follows:
\begin{eqnarray}\nonumber
\sin\tilde{\theta}_{13}&= &\left(\frac{\Delta_{31}}{\Delta_{31}-2aL}\right)\sin\theta_{13}\\\nonumber
\sin\tilde{\theta}_{23}&= &\sin\theta_{23}\\\nonumber
\sin\tilde{\theta}_{12}&= &-\frac{1}{2}\left(\frac{\Delta_{21}}{2aL}\right) \sin 2\theta_{12}\\\nonumber
\sin\tilde{\delta}& = &\sin\delta 
\label{EffectiveAngleFormula_NunokawaValleParke}
\end{eqnarray}
while the effective mass-squared differences are: 
\begin{eqnarray}\nonumber
\tilde{\Delta}_{21}&=& -2a L \\\nonumber
\tilde{\Delta}_{32}&=& \Delta_{31}-\Delta_{21}\cos^2\theta_{12}\\\nonumber
\tilde{\Delta}_{31}&=&\Delta_{31} -2a L
\label{EffectiveMassFormula}
\end{eqnarray}
where $\Delta_{kl}=\frac{\Delta m^2_{kl}L}{2E}$.


\providecommand{\href}[2]{#2}\begingroup\raggedright\endgroup

\end{document}